\documentclass[11pt,fleqn]{article}       
\usepackage{geometry}
\usepackage[T1]{fontenc}
\usepackage[utf8]{inputenc}
\usepackage{authblk}
\usepackage{mathrsfs}
\usepackage[font=small,labelfont=bf]{caption}
\usepackage{graphicx}
\usepackage{multirow}
\usepackage{amsmath,amssymb,amsfonts}
\usepackage{verbatim}
\usepackage{bm}
\usepackage{color}
\usepackage{ulem}
\usepackage{mathtools}
\usepackage{cite}
\usepackage{colortbl}
\definecolor{light-gray}{gray}{0.85}
\usepackage[pdftex,colorlinks,pdfpagelabels]{hyperref}
\hypersetup{
   bookmarksnumbered,
   citecolor={blue},
   linkcolor={blue},
   urlcolor={blue},
   filecolor={blue}
} 
\newcommand{\eVdist}{\kern-0.06em}

\newcommand{\kpc}{\:\text{kpc}}

\newcommand{\km}{\:\text{km}}
\newcommand{\pc}{\:\text{pc}}
\newcommand{\Mpc}{\:\text{Mpc}}

\newcommand{\s}{\:\text{s}}
\newcommand{\D}{\mathrm{d}}

\newcommand{\AddrTexas}{%
\textit{Department of Physics, The University of Texas at Austin, Austin, 78712 TX, USA}
}
\newcommand{\AddrStockholm}{
\textit{Oskar Klein Center for Cosmoparticle Physics, University of Stockholm, 10691 Stockholm, Sweden}
}
\newcommand{\AddrNordita}{
\textit{Nordita, KTH Royal Institute of Technology and Stockholm University, Roslagstullsbacken 23, 10691 Stockholm, Sweden}
}

\usepackage{fancyhdr}
 \usepackage[bottom]{footmisc}
\fancypagestyle{plain}{%
    \fancyhead[R]{UTTG-02-2021, NORDITA-2021-020}
    
}
\date{}
\title{\Large\bf Chain Early Dark Energy:\\Solving the Hubble Tension and Explaining Today's Dark Energy}
\author[1,2,3]{Katherine Freese\thanks{ktfreese@utexas.edu}}
\author[1,2]{Martin Wolfgang Winkler\thanks{martin.winkler@su.se}}
\affil[1]{\AddrTexas}
\affil[2]{\AddrStockholm}
\affil[3]{\AddrNordita}

\begin{document}
\maketitle
\vspace*{0mm}
\begin{abstract}
We propose a new model of Early Dark Energy (EDE) as a solution to the Hubble tension in cosmology, 
the apparent discrepancy between local measurements of the Hubble constant $H_0\simeq 74 \km\,\s^{-1}\Mpc^{-1}$
and $H_0\simeq 67 \km\,\s^{-1}\Mpc^{-1}$ inferred from the Cosmic Microwave Background (CMB).
In Chain EDE, the Universe undergoes a series of first order phase transitions, 
starting at a high energy vacuum in a potential, and tunneling down
 through a chain of every lower energy metastable minima.   
 As in all EDE models, the contribution of the vacuum energy 
 to the total energy density of the universe is initially negligible, but reaches $\sim 10\%$ around matter-radiation equality, before cosmological data require it to redshift away quickly -- at least as fast as radiation.
 We indeed obtain this required behavior with a series of $N$ tunneling events, and show that for $N>600$
 the phase transitions are rapid enough to allow fast percolation and thereby avoid large scale anisotropies in the CMB.  We construct a specific example of Chain EDE featuring a scalar field in
a quasiperiodic potential (a tilted cosine), which is ubiquitous in axion physics and, therefore, carries strong theoretical motivation.  Interestingly, the energy difference between vacua can be roughly the size of today's Dark Energy (meV scale). Therefore, the end result of Chain EDE could provide a natural explanation of Dark Energy, if the tunneling becomes extremely slow in the final step before the field reaches zero (or negative) energy. We discuss a simple mechanism which can stop the scalar field in the desired minimum. Thus Chain EDE offers the exciting prospect to explain EDE and Dark Energy by the same scalar field.
\end{abstract}
\clearpage

\section{Introduction}
The cosmological standard model $\Lambda$CDM has provided dramatic insights into the evolution of the universe, the formation of Large Scale Structure (LSS) and the physics of the Cosmic Microwave Background (CMB). However, one striking anomaly referred to as the `Hubble tension' has persisted over the last years and could indicate that $\Lambda$CDM is incomplete. Supernova redshift observations consistently prefer a value of $H_0\simeq 74 \km\,\s^{-1}\Mpc^{-1}$~\cite{Riess:2018byc,Riess:2019cxk} significantly different from $H_0\simeq 67 \km\,\s^{-1}\Mpc^{-1}$ as inferred from the CMB by Planck~\cite{Aghanim:2018eyx}. A possible resolution of this discrepancy requires an alteration of the expansion history compared to $\Lambda$CDM. In particular, if the sound horizon at matter-radiation decoupling is reduced, a larger value of $H_0$ would be derived from the CMB~\cite{Riess:2016jrr}.

The challenge is that any type of physics affecting the sound horizon induces further changes in the pattern of CMB fluctuations. Simple scenarios like an extra dark radiation increase the CMB-inferred Hubble parameter at the price of degrading the cosmological fit to the higher-$\ell$  temperature peaks and polarization of the CMB and baryon acoustic oscillation data~\cite{Bernal:2016gxb,Aghanim:2018eyx}. However, one particular form of energy dubbed `Early Dark Energy' (EDE) has been shown to resolve the Hubble tension while (so far) passing all cosmological tests~\cite{Karwal:2016vyq,Poulin:2018dzj,Poulin:2018cxd,Agrawal:2019lmo,Smith:2019ihp,Niedermann:2019olb,Hill:2020osr,Niedermann:2020dwg,Smith:2020rxx}. Initially, the energy density of EDE 
behaves roughly as a cosmological constant 
(${\rm log}(\rho_{\text{EDE}}) \sim$ constant)  
and its contribution to the total energy density of the universe is negligible. But as the universe cools down the ratio $\rho_{\text{EDE}}/\rho_{\text{tot}}$ grows and reaches $\sim 10\%$ around matter-radiation equality, before cosmological data require it to redshift away quickly -- at least as fast as radiation~\cite{Poulin:2018cxd}.

The original model of EDE involved a scalar field $\phi$ which is initially displaced from its minimum (see e.g.~\cite{Poulin:2018cxd,Agrawal:2019lmo,Smith:2019ihp}). As long as the Hubble friction dominates its evolution, the field value remains (nearly) constant and the corresponding potential energy behaves as a cosmological constant. But once the Hubble scale approaches the effective mass of the scalar field, $\phi$ becomes dynamical. While it performs coherent oscillations around its minimum, the energy density $\rho_\phi$ stored in $\phi$ decreases. A major drawback of the scalar field models is that $\rho_\phi$ typically redshifts away more slowly than radiation -- in contrast to what is required in the EDE scenario. Only for highly non-generic choices of the scalar field potential like $V\propto (1-\cos(\phi/f))^n$ with $n\geq 2$, a sufficiently fast decrease of the energy density can be realized~\cite{Poulin:2018cxd}.

An alternative model of EDE invokes a first order phase transition. The universe is trapped in a false minimum, but tunnels into the true minimum around matter radiation equality. Unfortunately, this scenario suffers from a shortcoming which bears resemblance to the `empty universe problem' ~\cite{Guth:1982pn} in old inflation~\cite{Guth:1980zm}: in order for EDE to stay around long enough the tunneling rate must be suppressed. This implies that during the phase transition bubbles of true vacuum are formed far away from each other. They grow to macroscopic sizes before eventually colliding with a neighboring bubble and releasing the energy contained in the bubble walls. As a consequence large scale anisotropies would arise and leave undesirable footprints in the CMB. 

In order to avoid these anisotropy problems in EDE, one can mimic solutions to the empty universe problem in Guth's original model of `old' inflation:
 (i) Double Field inflation~\cite{Adams:1990ds,Linde:1990gz}) employed an additional trigger field to make the tunneling rate time-dependent;
at first the tunneling rate is slow to allow for  a long enough period of inflation but then suddenly switches to rapid tunneling so that bubbles nucleate simultaneously throughout the Universe, collide, and reheat. Ref.~\cite{Niedermann:2019olb,Niedermann:2020dwg} employed the same approach, using a trigger field to obtain a
time-dependent tunneling rate, to construct a viable model of EDE.
 (ii) In Chain Inflation~\cite{Freese:2004vs,Freese:2005kt}, the Universe tunnels from a large vacuum energy through a series of minima of ever lower vacuum energy.  Each tunneling event is rapid enough for successful percolation and reheating, with hundreds of tunneling events required to add up to enough inflation. 
 
In this work we propose a new model of EDE in which the universe undergoes a series of phase transitions instead of just one. Drawing on its similarity with Chain Inflation~\cite{Freese:2004vs,Freese:2005kt}, we will dub our proposal `Chain Early Dark Energy' (Chain EDE). In the simplest case, Chain EDE is realized through a scalar field which starts at a vacuum with energy density $V_0 = \mathcal{O}( \text{eV}^4)$ and tunnels down through a chain of metastable minima of ever lower energy. The required quasiperiodic potential (e.g.\ a tilted cosine) is ubiquitous in axion physics and, therefore, carries strong theoretical motivation.
 With the appropriate choice of two parameters -- the initial EDE density $V_0$
  and the summed lifetime of vacua along the chain\footnote{As discussed further below, this summed lifetime is approximately given by $N/\Gamma^{1/4}$ where $N$ is the number of tunneling events and $\Gamma$ is the tunneling rate per phase transition.} -- the EDE fraction of the Universe's energy density is negligible at the beginning, raises to $\sim 10\%$ around matter-radiation equality and then decays away quickly, as shown in Figure~\ref{fig:edescenario} below. This is exactly the required behavior to resolve the Hubble tension.

Furthermore, in Section~\ref{sec:anisotropies}, we will argue that Chain EDE does not suffer from the anisotropy problem which plagues the single phase transition case. Given a series of metastable vacua, the requirement of EDE to survive until matter-radiation equality can be fulfilled even if each individual vacuum is short-lived. Hence, the bubbles of new vacuum in Chain EDE are created in close proximity and only grow for a short time before they percolate. We will show that the resulting anisotropies occur at small distance scales and do not spoil cosmological observables.

In Section~\ref{sec:hubble}, we will trace the evolution of the Chain EDE component over the history of the universe. In particular, we will show that the evolution is very similar as in the best fit EDE models~\cite{Smith:2019ihp} invoking oscillating scalar fields. Since the latter have been proven to resolve the Hubble tension via dedicated cosmological fits, the same can be claimed for Chain EDE.

Section~\ref{sec:model} presents a specific model realization of Chain EDE in form of a tilted cosine potential. We also introduce a mechanism for stopping the axion once the EDE is dissipated at $z\simeq 3300$, i.e.\ which prevents $\phi$ from further tunneling down the chain into the regime of negative vacuum energy.

Finally, in Section~\ref{sec:chainde} we will show that the energy difference between vacua in Chain EDE can naturally be of $\mathcal{O}(\text{meV})$. We will argue that if the tunneling field is still trapped in the lowest minimum with positive energy, it can account for the Dark Energy which dominates our universe today.

\section{Constraints on Phase Transitions from CMB Anisotropies}\label{sec:anisotropies}

In the first step we want derive constraints on one or several first order phase transition(s) that take place shortly before matter-radiation equality. A phase transition induces scale-dependent anisotropies which can spoil CMB and LSS observations.

\subsection{CMB Bounds on a Single Phase Transition}\label{sec:cmbsingle}

We consider a scalar field $\phi$ which is initially trapped in a false vacuum. The energy density of the scalar is assumed to be subdominant at all times such that the expansion of the universe is mainly controlled by the radiation component. Once the age of the universe approaches the life-time of the false vacuum, bubbles of true vacuum are formed with the energy of the false vacuum stored in the bubble walls. The bubble walls expand approximately at the speed of light and release their energy upon collision (e.g.\ into radiation). In order to avoid dangerous energy injection into the visible sector $\phi$ is assumed to be a dark sector field that does not couple directly to the visible sector.

If we denote the tunneling rate per volume  as $\Gamma$, we can determine the mean lifetime $\tau$ of the universe in the false vacuum by requiring $\Gamma \mathcal{V}_4(\tau)=1$. Here $\mathcal{V}_4(\tau)$ is the spacetime volume of the past lightcone at time $t=\tau$ (at an arbitrary position)\footnote{$\mathcal{V}_4(\tau) = \int_0^\tau {4 \pi \over 3} r^3(t) dt$
with $r(t) = a(t)\int_t^\tau {dt' \over a(t')}$, where $a$ denotes the scale factor of the universe.}. For a radiation-dominated universe we obtain
\begin{equation}\label{eq:tau}
\tau=\left(\frac{105}{8\pi\,\Gamma}\right)^{1/4}\simeq 1.4 \times \Gamma^{-1/4}\,.
\end{equation}
We can invert this equation and express $\Gamma$ in terms of the redshift of the phase transition $z_b\equiv z(\tau)$,
\begin{equation}\label{eq:redshiftphase}
\Gamma^{-1/4}\simeq 4.4\times 10^4\text{yr}\times \left(\frac{3500}{z_b}\right)^2 \simeq 13.5\:\text{kpc}\times\left(\frac{3500}{z_b}\right)^2\,.
\end{equation}
One can easily verify that the typical distance between the bubble nucleation centers is also given by $\sim \Gamma^{-1/4}$. It is convenient to consider the comoving bubble separation
\begin{equation}\label{eq:bubblesep}
d_b \simeq \Gamma^{-1/4} z_b \simeq 47\:\text{Mpc}\times\left(\frac{3500}{z_b}\right)\,.
\end{equation} 
This is because $d_b$ will also correspond to the comoving size of the bubbles upon percolation (assuming that bubble walls expand at the speed of light until collision). After the bubble walls transferred their energy, anisotropies of comoving size $d_b$ are present in the dark sector which are gravitationally transferred to the visible sector. These anisotropies can leave imprints in the CMB which are not observed in the data~\cite{Niedermann:2019olb,Niedermann:2020dwg}. Indeed if the energy density
in the vacuum is 10\% of the radiation, one might expect ${\delta \rho / \rho} \sim 10\%$ at the time of the phase transition, growing up to $\mathcal{O}(1)$ at the time of the CMB epoch. In the following we will require the anisotropies to occur at small enough scales to be unobservable in CMB and LSS data. 

The anisotropies occur at an angle in the sky which can be estimated as follows~\cite{Niedermann:2020dwg},
\begin{equation}\label{eq:theta}
\theta \simeq \frac{d_b}{D_{\text{CMB}}}\,,
\end{equation}
where $D_{\text{CMB}}$ denotes the comoving angular diameter distance of the CMB
\begin{equation}\label{eq:diameterdistance}
D_{\text{CMB}} = \int\limits_0^{z_{\text{CMB}}} dz\, H^{-1}(z)\simeq 14\:\text{Gpc}\,.
\end{equation}
Current observations of the CMB temperature power spectrum by Planck and earth-bound detectors reach up to $\ell\simeq 4000$~\cite{Aghanim:2018eyx}. Even smaller scales are accessible through the Lyman-$\alpha$ forest which covers the dynamical range $k\simeq (0.1-10)\:h\,\text{Mpc}^{-1}$~\cite{Zaroubi:2005xx} corresponding to $\ell\simeq 10^3 -10^5$~\cite{Akrami:2018odb}. If we use the stronger Lyman-$\alpha$ constraint we need to require 
\begin{equation}\label{eq:thetamax}
\theta < \frac{\pi}{\ell_\text{max}} = 0.002^\circ
\end{equation}
on the angular scale of anisotropies.  We use Eq.~\eqref{eq:diameterdistance},~\eqref{eq:thetamax} in Eq.~\eqref{eq:theta} to find a bound on $d_b$, which using Eq.~\eqref{eq:bubblesep} translates to
\begin{equation}\label{eq:CMBGamma}
 \Gamma^{-1/4} < 0.14\kpc\times\frac{3500}{z_b}\,.
\end{equation}
Combining Eq.~\eqref{eq:redshiftphase} and~\eqref{eq:CMBGamma} we obtain a constraint on the redshift of the phase transition
\begin{equation}\label{eq:zconstraint}
z_b>3.3\times 10^5\,.
\end{equation}
However, in order to resolve the Hubble tension EDE has to stay around until redshift $z\sim 3500$~\cite{Poulin:2018cxd,Smith:2019ihp}. Therefore, EDE models with a single phase transition suffer from unacceptable anisotropies in the CMB.

As mentioned in the introduction, a caveat to this argument which employs a time-dependent tunneling rate has been pointed out in~\cite{Niedermann:2019olb,Niedermann:2020dwg}: if $\Gamma$ is initially small, but increases around matter-radiation equality the lifetime of the universe in the false vacuum and the duration of the phase transition (which controls the size of anisotropies) can effectively be decoupled. In particular, the anisotropies can be made compatible with CMB constraints if the increase in $\Gamma$ occurs rapidly. The required mechanism, which includes the EDE field and an additional trigger field, has first been introduced in the context of double field inflation~\cite{Adams:1990ds,Linde:1990gz}.

\subsection{CMB Bounds on a Series of Phase Transitions}\label{sec:cmbchain}

\begin{figure}[t]
\begin{center}
\includegraphics[width=0.9\textwidth]{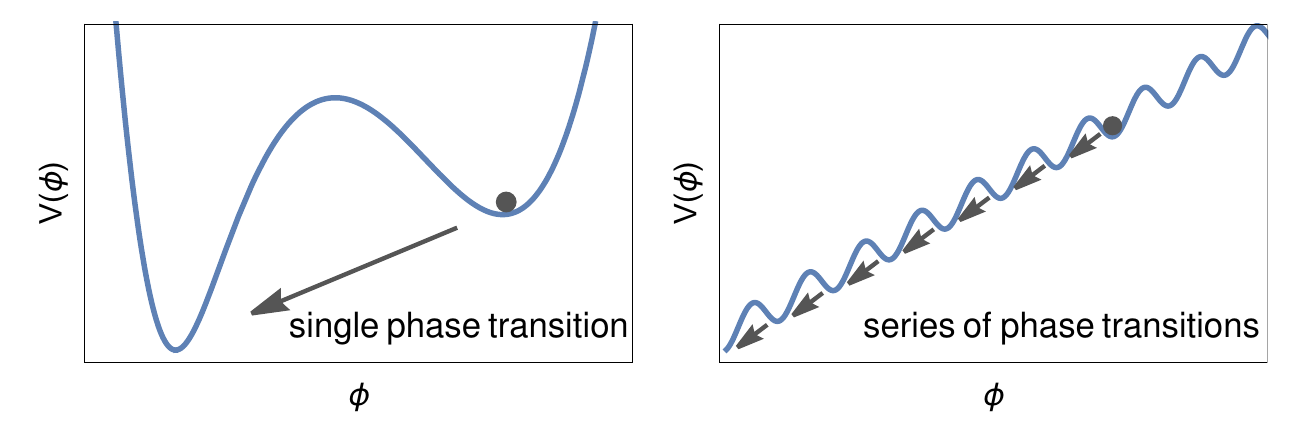}
\end{center}
\vspace{-3mm}
\caption{Models of EDE in which the energy is initially stored in a scalar field. In the left panel the EDE is dissipated through a single phase transition. In the right panel a series of phase transitions occur. We denote the scenario in the right panel as Chain EDE.}
\label{fig:phase}
\end{figure}

We now turn to an alternative possibility to evade the CMB constraints. 
In Chain EDE the vacuum energy is dissipated through a series of phase transitions instead of just one. As a simple realization we consider a scalar field which tunnels along a chain of false vacua with decreasing energy (see Fig.~\ref{fig:phase} for illustration). In order to keep the discussion simple, we assume that the tunneling rate per volume $\Gamma$ remains constant for all vacuum transitions within the chain.  Again, the energy density of the scalar is assumed to be subdominant at all times such that the expansion of the universe is mainly controlled by the radiation component. 

Initially, the field sits in its highest energy minimum; in the meantime the (dominant) radiation density and the Hubble parameter decrease with the expansion of the Universe. Once the Hubble parameter drops to the point where it becomes comparable to the tunneling rate, $\Gamma/H^4 \sim 1$, the first phase transition in the chain takes place. 
 The first transition can be treated completely analogously to the single phase transition case in the previous section.
Thus as in the case of the single phase transition, it occurs when the lifetime of the universe reaches $1.4 \times \Gamma^{-1/4}$  (cf.~\eqref{eq:tau}). 

 Since bubbles are nucleated at every tunneling step, we need to apply the CMB constraints to each of the vacuum transitions. 
 We begin by finding the most restrictive bounds, in which we follow the same logic as in Section~\ref{sec:cmbsingle} for a single phase transition, and require the first phase transition to happen on scales not observable in the CMB. These restrictive bounds require a large number of phase transitions ($N>20,000$). However, we then reexamine the bounds for Chain EDE and argue that they likely do not need to be so restrictive, since the amplitude of CMB anisotropies in each of the phase transitions is reduced compared to the single phase transition case. Thus the number of phase transitions must instead satisfy $N>600$.
 
  \subsubsection{Most restrictive bounds, modeled on single phase transition case:}
First we find the most restrictive possible bounds on a series of phase transitions, by requiring the CMB anisotropies to occur on small enough scales to be unobservable in the CMB. Eq.~\eqref{eq:CMBGamma} implies that the strongest CMB constraints always apply to the first phase transition.
This can easily be understood since the corresponding anisotropies have the longest time available to grow by the expansion of the universe.  To be unobservable in the CMB,
the first phase transition, again as found in the case of a single phase transition,  would need to occur at redshift $z> 3.3 \times 10^5$ (see Eq.~\eqref{eq:zconstraint}), 
which according to Eq.~\eqref{eq:redshiftphase} translates to 
a  tunneling time 
\begin{equation}\label{eq:gamma14limit}
\Gamma^{-1/4} < 5\:\text{yr}\simeq 1.5\pc\,
\end{equation} 
(comparable to the age of the Universe at that redshift).

In the case of many vacua each tunneling step reduces the EDE only by a small amount. If there are sufficiently many transitions $N$ along the chain the EDE can stay around until matter-radiation equality while still satisfying the constraint above.
In order to find the minimal number $N$ we need to determine the (mean) lifetime of the universe in each vacuum. 

We realize that for each subsequent transition the lifetime per vacuum decreases.  Although we have assumed constant $\Gamma$
 for all the transitions, the Hubble parameter decreases with redshift during radiation domination as $H \propto z^2$.
 Thus the ratio $\Gamma/H^4 \propto z^{-8}$ increases rapidly, so that after the first few phase transitions
 the field has reached the fast tunneling regime $\Gamma/H^4 \gg 1$.
The number of vacuum transitions per time is obtained by counting the bubble walls hitting an observer at a fixed point in space. The first transition occurs when the lifetime of the universe, corresponding to $1/(2H)$ reaches $1.4 \times \Gamma^{-1/4}$ (cf.~\eqref{eq:tau}). Therefore, the vacuum bubbles seeded by the first transition are generated at a distance $\Gamma^{-1/4}=\mathcal{O}(1/H)$ from each other and take around one Hubble time to collide. Their evolution is, hence, significantly affected by the expansion of the universe. For subsequent transitions with $\Gamma/H^4\gg 1$, a large number of bubbles is created per Hubble four-volume implying that the bubbles collide before they `realize' the expansion.

Simulations of colliding bubbles have recently been performed in~\cite{Winkler:2020ape}. Even though the simulations assumed an inflationary background the results in the limit $\Gamma/H^4\gg 1$ also apply to radiation domination since the bubbles are not affected by the expansion as we argued. For $\Gamma/H^4\gg 1$ we extract the time of the universe in one vacuum $i$ as
\begin{equation}\label{eq:simulation}
\tau_i = \frac{\Delta\phi}{d\langle\phi\rangle/dt} \simeq 0.7\;\Gamma^{-1/4}\,,
\end{equation}
where $\Delta\phi$ is the field-space distance between minima. Comparing~\eqref{eq:tau} for the first phase transition and~\eqref{eq:simulation} for later phase transitions in the series, we see that $\tau_i$ decreases by an $\mathcal{O}(1)$-factor along the chain. 

In practice we will apply~\eqref{eq:simulation} to all vacuum transitions. Since $\Gamma/H^4\gg 1$ is satisfied after the first few tunneling steps, the error we make by this assumption is negligible. The total time $\tau$ elapsed after $N$ phase transitions can, hence, be estimated as
\begin{equation}
\tau =N \tau_i = 0.7\,N\times \Gamma^{-1/4} < N\times 1.1\pc\,.
\end{equation}
In the last step we applied the constraint derived from CMB anisotropies~\eqref{eq:gamma14limit}. Requiring that EDE remains present until redshift $z\sim 3500$ leads to the following constraint on the number of vacuum transitions
\begin{equation}\label{eq:nconstraint1}
N > 2 \times 10^{4}\,.
\end{equation}
We emphasize that we derived this constraint by requiring that anisotropies occur on small scales which are experimentally inaccessible. 

\subsubsection{Less Restrictive Bounds since there are Multiple Phase Transitions in the Chain}
The bound in Eq.(\ref{eq:nconstraint1}) is  unduly restrictive due to the fact that there are multiple
phase transitions.  The energy dissipation through a series of phase transitions (compared to just one) suppresses the amplitude of EDE-induced CMB anisotropies by a factor $\sim 1/N$ so that the anisotropies induced by Chain EDE could be acceptable even if they fall into the observable range of scales. While a more dedicated analysis goes beyond the scope of this work, one would expect the density perturbations sourced by EDE at redshift $z$ to be bounded by
\begin{equation}
\frac{\Delta\rho_{\text{EDE}}}{\rho_{\text{tot}}}\ \lesssim \frac{1}{N}\frac{\rho_{\text{EDE}}(z)}{\rho_{\text{tot}}(z)}\,,
\end{equation}
where $\rho_{\text{EDE}}(z)/\rho_{\text{tot}}(z)$ is the fraction of the total energy density of the universe contained in EDE at redshift $z$. Assuming $\rho_{\text{EDE}}/\rho_{\text{tot}} \sim 0.1$ at $z\sim 3500$ in order to resolve the Hubble tension and logarithmic (linear) growth of perturbations during radiation (matter) domination, we can determine the maximal amplitude of EDE-induced fluctuations as a function of scale and redshift. Requiring $\Delta\rho_{\text{EDE}}/\rho_{\text{tot}} < 10^{-5}$ at last scattering for scales observable in the CMB and in LSS then leads to the constraint
\begin{equation}
 N > 6\times 10^2\,,
\end{equation}
which is significantly weaker than~\eqref{eq:nconstraint1}. 

We conclude that if EDE disappears not by a single phase transition, but instead by a series of phase transitions, dangerous anisotropies can be avoided. The advantage of multiple phase transitions is two-fold: (i) the amplitude of the anisotropies is reduced by the number of transitions $N$; (ii) the scale of the anisotropies is smaller.  For the case of one phase transition, requiring the EDE to stick around long enough leads to a large bubble size and large scale anisotropies; for the case of many transitions, each phase transition need only last $1/N$ of the total EDE epoch and the faster tunneling rate produces smaller bubbles (at percolation) and smaller scale anisotropies.  Requiring at least $\mathcal{O}(10^4)$ transitions one is definitely on the safe side since anisotropies only occur at scales which are experimentally inaccessible. Most likely, even a few hundred transitions are sufficient since the amplitude of EDE-induced fluctuations is strongly suppressed in this case.

\section{Solution to the Hubble Tension} \label{sec:hubble}

The Hubble tension consists in the apparent discrepancy between local measurements of the Hubble constant yielding $H_0\simeq 74 \km\,\s^{-1}\Mpc^{-1}$~\cite{Riess:2018byc,Riess:2019cxk} and $H_0\simeq 67 \km\,\s^{-1}\Mpc^{-1}$ inferred from the CMB~\cite{Aghanim:2018eyx}. EDE resolves the discrepancy by adding an additional energy component which reduces the sound horizon $r_s$ at recombination. In order to preserve the angular size of the first peak in the CMB, the decrease of $r_s$ needs to be compensated by a reduction of the angular diameter distance of the CMB $D_{\text{CMB}}$. As apparent from~\eqref{eq:diameterdistance} this in turn leads to an increase of $H_0$ inferred from the CMB compared to $\Lambda$CDM, i.e.\ to a resolution of the Hubble tension. 

Since a shorter sound horizon would also shift the position of the higher CMB peaks to larger $\ell$ and affect their amplitude, further cosmological parameters including the dark matter density, the baryon density and the scalar spectral index need to be modified in order to balance this effect~\cite{Poulin:2018cxd,Agrawal:2019lmo,Smith:2019ihp}. Furthermore, to minimize the impact on other successful $\Lambda$CDM predictions, EDE should redshift away at least as fast as radiation at $z\lesssim 3000$.

We begin by discussing previously proposed models of EDE and then turn to our model of Chain EDE.

\subsection{Oscillating Scalar Field}\label{sec:oscillating}

Previous models of EDE invoke a scalar field $\phi$ in a dark sector which exhibits a potential of the form~\cite{Poulin:2018cxd,Smith:2019ihp}
\begin{equation}\label{eq:standardede}
 V = m^2 f^2 \left[ 1- \cos\left(\frac{\phi}{f}\right)\right]^n\,,
\end{equation}
where $m$ and $f$ are parameters of mass dimension one and $n$ is an integer number. Other choices, e.g.\ $V\propto \phi^{2n}$, have also been considered~\cite{Agrawal:2019lmo}. One can easily trace the time evolution of the EDE energy density $\rho_{\text{EDE}}= \dot{\phi}^2/2 + V$ by solving the homogeneous Klein-Gordon equation
\begin{equation}
 \ddot{\phi} + 3 H \dot{\phi} + V^\prime(\phi) = 0\,.
\end{equation}
The scalar field is initially frozen at the field value $\phi_0$ by the Hubble friction and $\rho_{\text{EDE}}$ remains constant. Once $H$ falls below the the effective mass $V''(\phi_0)$ the field starts to perform damped oscillations around its minimum. During this period, the energy oscillates around the asymptotic solution
\begin{equation}
\rho_{\text{EDE}} \propto a^{-6n/(n+1)}\,.
\end{equation}
where $a$ denotes the scale factor of the universe. Since, during the oscillation period, $\rho_{\text{EDE}}$ must redshift at least as fast as radiation $n\geq 2$ is required. 

In~\cite{Smith:2019ihp} the cosmological predictions of the EDE model were investigated in a combined fit to the Planck power spectra~\cite{Aghanim:2015xee}, BAO data~\cite{Beutler:2011hx,Ross:2014qpa,Alam:2016hwk} and supernova measurements of the Hubble constant~\cite{Scolnic:2017caz,Riess:2019cxk}. It was found that the cases $n=2$ ($n=3$) reduce the total $\chi^2$ by 16 (20) compared to $\Lambda$CDM -- suggesting a clear preference for the EDE component. In Fig.~\ref{fig:comparison} we depict the evolution of $\rho_{\text{EDE}}$ for the best fit point with $n=2$. As can be seen, the Early Dark Energy fraction reaches a maximum of $\rho_{\text{EDE}}/\rho_{\text{tot}}=0.09$ at $z=3111$. The best fit points features a Hubble constant $H_0\simeq 71.6 \km\,\s^{-1}\Mpc^{-1}$ close to the value preferred by local measurements.

\subsection{Chain EDE}

In the following we want to argue that Chain EDE is capable of resolving the Hubble tension. We have seen that the EDE solution requires $\rho_{\text{EDE}}$ to contribute significantly ($\sim 10\%$) around matter-radiation equality, but then to disappear quickly. However, the cosmological fit is expected to be insensitive to the details of the underlying model. In fact, a simple modeling of the EDE component in an effective fluid approach~\cite{Poulin:2018cxd} yielded very similar results compared to the full implementation of the oscillating scalar field model~\cite{Smith:2019ihp}. In this light we can refrain from performing a full cosmological fit for Chain EDE. Instead, we will show that Chain EDE is able to closely reproduce the redshift-dependence of $\rho_{\text{EDE}}(z)$ in the oscillating scalar field model.

In Chain EDE, the energy density of the EDE sector $\rho_{\text{EDE}}=\rho_\phi+\rho_{\text{wall}}+\rho_{\text{DR}}$ consists of three components, namely
\begin{itemize}
\item the vacuum energy stored in the scalar field $\rho_\phi$,
\item the energy of bubble walls $\rho_{\text{wall}}$,
\item the energy density $\rho_{\text{DS}}$ created by the collision of bubble walls which may consist of dark radiation, gravity waves and anisotropic stress~\cite{Niedermann:2019olb,Niedermann:2020dwg}. We will refer to this component as the energy density of the dark sector (DS) in the following.
\end{itemize}
Let us first investigate how the potential energy $\rho_\phi$ evolves. Initially, all energy of the EDE sector is stored in $\phi$ and, hence, $\rho_{\text{EDE}}=\rho_\phi\equiv V_0$. With each tunneling process the potential energy is reduced by the energy difference between vacua $\Delta V=V_0/N$, where $N$ again denotes the number of transitions required to dissipate the energy in $\phi$. To keep the discussion simple, we take both $\Delta V$ and the decay rate (per volume) $\Gamma$ to be constant along the entire chain of vacua.

\begin{figure}[t]
\begin{center}
\includegraphics[width=1\textwidth]{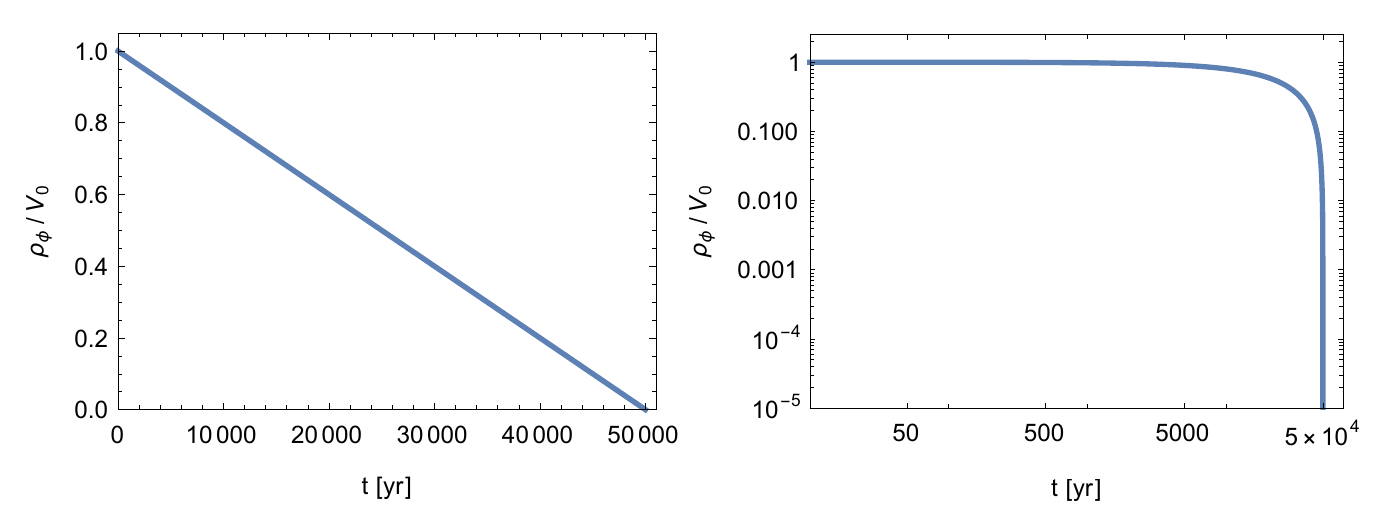}
\end{center}
\vspace{-3mm}
\caption{Time-evolution of the energy density $\rho_\phi$ in Chain EDE (in units of the initial energy density $V_0$). The same $\rho_\phi$ is depicted on a linear scale (left panel) and on a log-log-scale (right panel). The figure illustrates that a linear decrease of $\rho_\phi$ looks almost like a step-function in a log-log plot. We note that the individual phase transitions in the chain are too rapid to be visible in the plot which hence looks like a line. Notice that redshift moves from right to left in this figure since we are plotting time on the x-axis --- the reverse of Figures~\ref{fig:edescenario} and~\ref{fig:comparison} below.
}
\label{fig:linear}
\end{figure}

If there are sufficiently many vacua in the chain, we can approximate $\rho_\phi$ to continuously decrease with
\begin{equation}\label{eq:lintime}
\dot{\rho}_\phi\simeq - \frac{\Delta V}{\tau_i}\,,
\end{equation}
with $\tau_i$, denoting the time spent in one vacuum as given in~\eqref{eq:simulation}. 

Eq.~\eqref{eq:lintime} implies that $\rho_\phi$ decreases linearly with time. Naively, this seems at odds with the requirements that EDE should initially behave as a cosmological constant and then disappear quickly. However, the intuition fails here, since these requirements need to be imposed on a logarithmic scale. And, in fact, a linearly decreasing function $\rho_\phi$ looks almost like a step function in a log-log-plot as we show in Fig.~\ref{fig:linear}.

Furthermore, since (most) transitions in Chain EDE occur quickly (compared to the Hubble time), we will approximate the energy transfer from $\rho_\phi$ to $\rho_{\text{DS}}$ as instantaneous and neglect $\rho_{\text{wall}}$ in the following.

The dark sector is permanently heated by vacuum transitions of $\phi$. At the same time the energy density of $\rho_{\text{DS}}$ redshifts with the scale factor as $a^{-3(1+w)}$, where $w$ stands for the equation-of-state parameter. The evolution of the energy densities in the EDE sector is, hence, governed by the following set of differential equations
\begin{align}\label{eq:diffeq}
(1+z)\,\frac{\D\rho_\phi}{\D z}&\simeq \frac{1.4\;\Delta V\; \Gamma^{1/4}}{H(z)}\,,\nonumber\\
(1+z)\,\frac{\D\rho_{\text{DS}}}{\D z}&\simeq -\frac{1.4\;\Delta V\; \Gamma^{1/4}}{H(z)} + 3(1+w)\,\rho_{\text{DS}}\,,
\end{align}
where we traded the time-dependence for a redshift dependence and used Eq.(\ref{eq:simulation}). 
Note that in the 2nd of the above equations, the first term on the right hand side (RHS) is simply the negative of the RHS  of the 1st equation (as energy is transferred from the vacuum to the dark sector), and the second term is the redshifting of the DS component.

The equation-of-state parameter depends on the underlying dissipation mechanism for the vacuum energy. It is expected that bubble walls release their energy into small scale anisotropic stress, dark radiation and, subdominantly, gravity waves. The distribution among these forms of energy is, unfortunately, highly model-dependent. The generation of dark radiation e.g. requires the availability of light final states in the dark sector. Furthermore, while $w=1/3$ for dark radiation, the equation-of-state parameter for anisotropic stress is not precisely known. A number of heuristic arguments suggest that it falls in the range $1/3 < w <1$~\cite{Niedermann:2020dwg}. Luckily, the EDE solution to the Hubble tension only requires $w\geq 1/3$~\cite{Poulin:2018cxd} which is satisfied by all forms of energy emerging from the bubble collisions. We can, therefore, refrain from a more detailed investigation of the equation-of-state parameter and simply consider the limiting cases $w=1/3$ and $w=1$ in order to bracket the uncertainties.

The evolution of the EDE sector is coupled to the visible sector through the expansion rate. We will later perform a full numerical solution of~\eqref{eq:diffeq} taking into account the impact of EDE on $H$. However, in order to roughly understand how $\rho_{\text{EDE}}$ evolves with redshift, it is instructive to perform an analytic estimate which neglects the (subdominant) impact of EDE on the Hubble parameter. Since the phase transitions occur during radiation domination we can then approximate $H(z)\simeq 0.25\,z^2\times 10^{-8}\,\kpc^{-1}$. Solving the first equation in Eq. (\ref{eq:diffeq}), we obtain
\begin{equation}\label{eq:rhophi}
 \rho_\phi = V_0 \left( 1 - 2.8\times 10^8\kpc\;\frac{\Gamma^{1/4}}{N\,z^2}\right)\,,
\end{equation}
where we replaced $V_0/\Delta V$ by the total number of transitions $N$. 
We note that the change in the potential scales as $1/z^2$ i.e. drops off linearly with time, as expected.
The expression above holds as long as $\rho_\phi\geq 0$. We assume that $\phi$ settles in a stable minimum at $V \sim 0$ once the entire potential energy has been dissipated\footnote{We  assume tunneling into anti-de Sitter does not happen.}. In order to resolve the Hubble tension this should occur at $z_*\simeq 3000$ which allows us to constrain $\Gamma$,
\begin{equation}
\frac{\Gamma^{1/4}}{N}\simeq 0.03 \kpc^{-1} \left(\frac{z_*}{3000}\right)^2\,.
\end{equation}
Notice that, in terms of the background evolution, models of Chain EDE are indistinguishable as long as they feature the same $\Gamma^{1/4}/N$. The absence of dangerous CMB anisotropies requires $N \gtrsim 10^4$ transitions (see Sec.~\ref{sec:cmbchain}), but one is otherwise free to choose $N$. Using~\eqref{eq:rhophi} the differential equation for the dark sector energy density can also be solved analytically such that we arrive at
\begin{align}\label{eq:analytic}
    \rho_\phi &= V_0 \begin{cases}
    1 - \left(\frac{z_*}{z}\right)^2 & \;\;z>z_* \\
    0                            & \;\;z<z_*                       \end{cases}
\;,\nonumber\\[1mm]
    \rho_{\text{DS}}&= \frac{2\,V_0 }{5+3w}\begin{cases}\left(\frac{z_*}{z}\right)^2  & \;\;z>z_* 
    \\[1mm]
    \left( \frac{z}{z_*}\right)^{3+w} & \;\;z<z_* 
\end{cases}
\;,\nonumber\\[1mm]
\rho_{\text{EDE}}&=\rho_\phi+ \rho_{\text{DS}}=V_0 \begin{cases}
   1 - \frac{3+3w}{5+3w}\left(\frac{z_*}{z}\right)^2 & \;\;z> z_* \\[1mm]
    \frac{2 }{5+3w}\left( \frac{z}{z_*}\right)^{3+w}  & \;\;z<z_* 
\end{cases}\;.
\end{align}         
Note that we have defined the EDE component of the Universe $\rho_{\text{EDE}}$ to include both the chain vacuum energy $\rho_\phi$ plus the dark sector $\rho_{\text{DS}}$ that it decays into.
We thus find that $\rho_{\text{EDE}}$ behaves approximately as a cosmological constant until $z_*$ and then redshifts away at least as fast as radiation (since $w\geq 1/3$). Hence, Chain EDE meets the criteria for a successful solution to the Hubble tension.

\begin{figure}[t!]
\begin{center}
\includegraphics[width=0.7\textwidth]{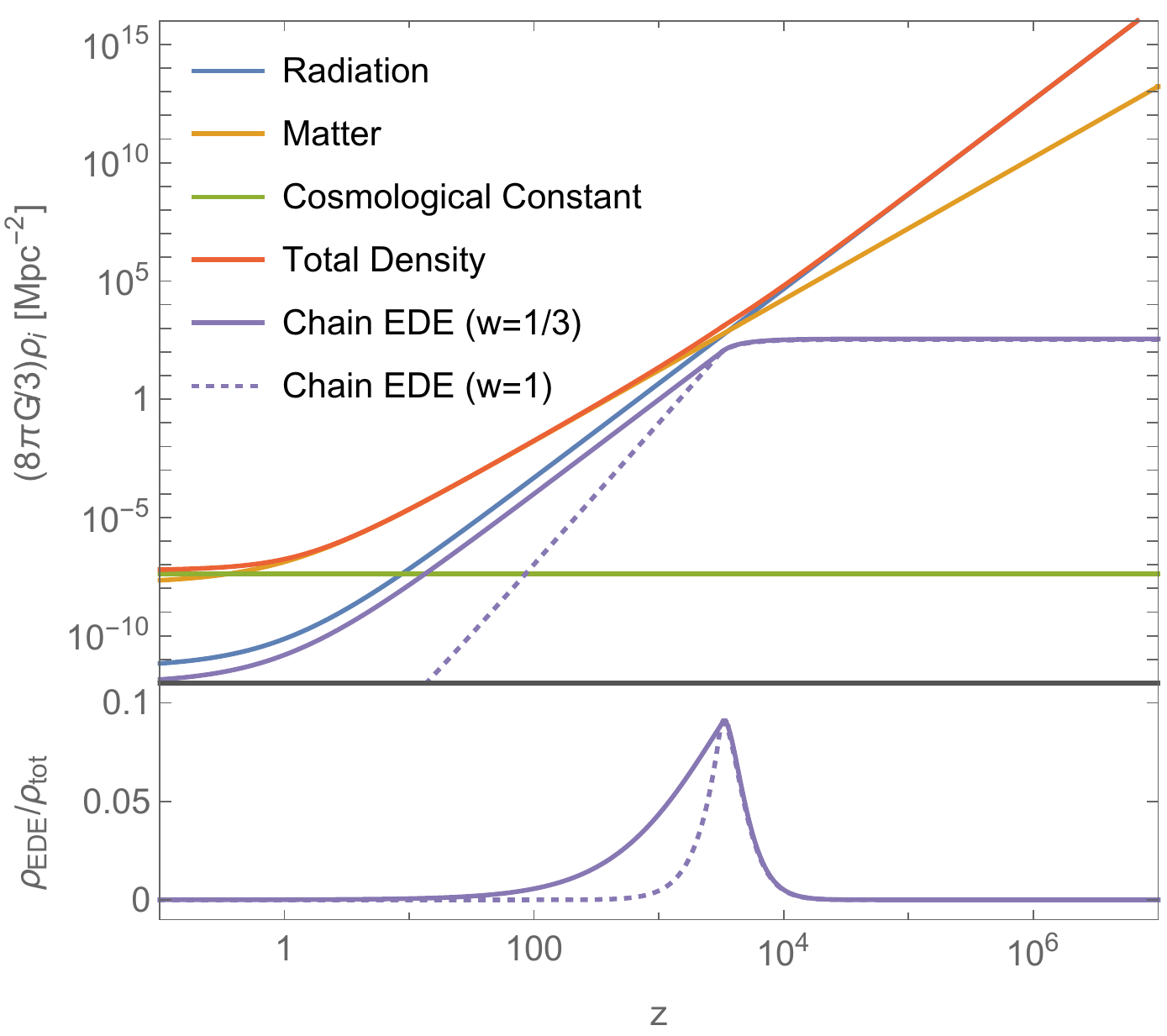}
\end{center}
\caption{Evolution of the radiation, matter, Dark Energy and EDE densities in the Chain EDE scenario. The cases $w=1$ and $w=1/3$ bracket the uncertainties in the equation-of-state parameter of the final state effective fluid generated by vacuum transitions in the EDE sector. 
}
\label{fig:edescenario}
\end{figure}

We have then performed a numerical solution of~\eqref{eq:diffeq} taking into account all subdominant contributions to the Hubble parameter and the full dynamics of the visible sector. The initial vacuum energy and the decay rate were chosen as $V_0=0.25\:\text{eV}^4$ and $\Gamma^{1/4}/N = 0.04\kpc^{-1}$. In Fig.~\ref{fig:edescenario} we depict the resulting evolution of the energy densities in radiation, matter, Dark Energy and EDE (the latter essentially follows our analytic estimate~\eqref{eq:analytic}). The cases $w=1/3$ and $w=1$ are depicted separately. As can be seen, the EDE component amounts to an energy injection strongly peaked around matter-radiation equality, while it plays virtually no role outside this window.

In Fig.~\ref{fig:edescenario}, we have plotted $\rho_{\text{EDE}}=\rho_\phi+ \rho_{\text{DS}}$;
i.e. the EDE curve has contributions both from the chain vacuum energy and the dark sector it decays into.
If we were to plot only the vacuum component $\rho_\phi(z)$, (the first equation in Eq.(\ref{eq:analytic})), in this log-log plot it would look similar to a step function: essentially flat for all $z>z_*$ and plummeting to $\rho_\phi$=0 at $z=z_*$.   As a reminder, for sufficiently many vacua, we can treat $\rho_\phi$ as continuously decreasing for the purposes of these figures.  Below $z_*$ the vacuum energy has converted to $\rho_{\text{DS}}$ which redshifts away.

\begin{figure}[t]
\begin{center}
\includegraphics[width=0.7\textwidth]{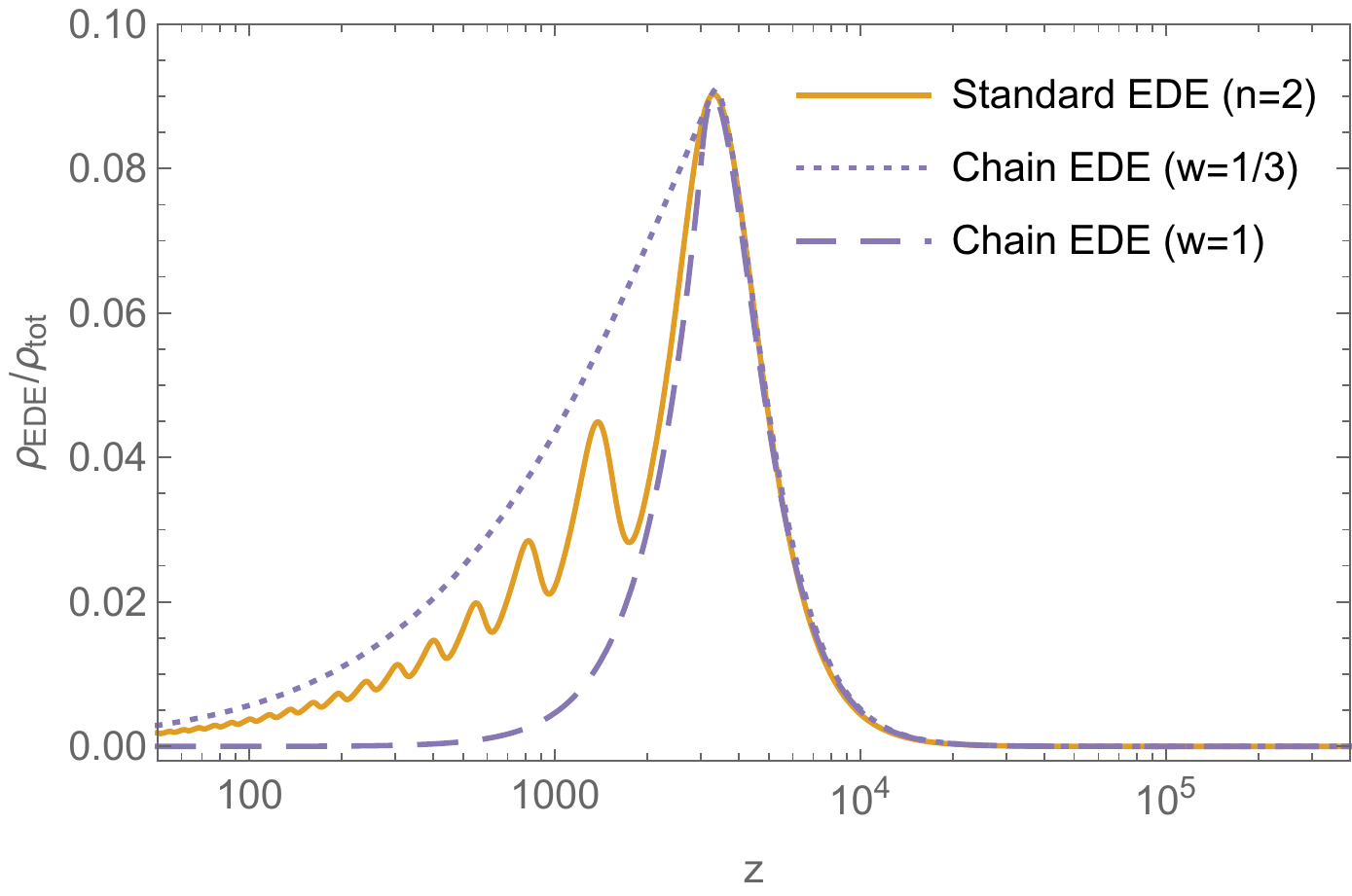}
\end{center}
\caption{Energy density in the EDE component compared to the total energy density of the universe. The orange line shows the best fit EDE solution for an oscillating scalar field ($n=2$ model from~\cite{Smith:2019ihp}). The two purple lines refer to Chain EDE for the parameters stated in the text. The cases $w=1$ and $w=1/3$ bracket the uncertainties in the equation-of-state parameter of the final state effective fluid generated by the bubble wall collisions. 
}
\label{fig:comparison}
\end{figure}

In Fig.~\ref{fig:comparison} we compare the evolution of $\rho_{\text{EDE}}$ in the Chain EDE scenario and in the oscillating scalar field model~\cite{Smith:2019ihp} described in the previous section (which we refer to as standard EDE in the figure). For the latter we have chosen the best fit point with $n=2$.

We observe that all three cases in Fig.~\ref{fig:comparison} are virtually indistinguishable for $z>z_*$. For $z<z_*$ the Early Dark Energy density in the standard EDE scenario oscillates between the Chain EDE solutions with $w=1/3$ and $w=1$. Since the true equation-of-state parameter of Chain EDE is expected to lie between the two extremes, a very similar scaling of $\rho_{\text{EDE}}$ in Chain EDE and in the standard EDE scenario is expected.

An explicit proof that Chain EDE resolves the $H_0$-problem of $\Lambda$CDM would require a dedicated cosmological fit including the full modeling of the EDE component at the fluctuation level. However, we have shown that Chain EDE follows almost exactly the background evolution of the standard EDE scenario. Since the latter has been proven to resolve the Hubble tension, we consider it almost certain that the same is true for Chain EDE.

As discussed in the introduction, we must fix two parameters to obtain successful Chain EDE.
Phenomenologically, we must obtain the correct values for 
$z_* \sim 3500$ and $\rho_{EDE}(z_*) \sim 0.1 \rho_{tot}(z_*)$.
As shown in Eqs. (20,21) above, 
 these two requirements can be achieved by the choice of the two parameters $V_0$ and tunneling rate $\Gamma^{1/4}/N$.
 In the next section, for the particular case of a tilted cosine potential, we will explicitly make choices for these two parameters that lead to successful Chain EDE.

\section{Model Realization of Chain EDE}\label{sec:model}

As a simple realization of Chain EDE, we consider an axion field $\phi$ in a quasi-periodic potential
\begin{equation}\label{eq:potential}
  V(\phi)=-\mu^3\phi  + \Lambda^4 \cos\left(\frac{\phi}{f} \right) +V_0 \,,
\end{equation}
where $f$ denotes the axion decay constant, while the parameters $\mu$ and $\Lambda$ control the strength of the shift symmetry breaking and the barrier height of individual minima. Finally, $V_0$ stands for a possible constant in the potential. Without loss of generality we can take $\phi_0\simeq 0$ as the initial field value such that $V_0$ corresponds to the initial EDE energy density (thus matching our previous definition of $V_0$). For convenience, we define the parameter $x = f\mu^3/\Lambda^4$. Given that $x<1$, the potential features an (infinite) series of minima with decreasing vacuum energy. 

The tunneling rate $\Gamma$ between two minima is given by
\begin{equation}\label{eq:gamma}
  \Gamma = A \, e^{-S_E}\,,
\end{equation}
where $S_E$ stands for the Euclidean action of the bounce solution~\cite{Coleman:1977py}, while $A$ denotes a prefactor which incorporates quantum fluctuations about the classical action~\cite{Callan:1977pt}. In a recent paper~\cite{Winkler:2020ape}, we provided new analytic estimates for the bounce action and the tunneling rate in periodic potentials which replace the thin-wall approximation in the regime of fast tunneling.
Specifically, we derived the following analytic approximation of the tunneling rate for the potential in Eq.~\eqref{eq:potential},
\begin{equation}\label{eq:final_approximation}
\Gamma \simeq \frac{\Lambda^8}{f^4}\,(1-x^2)\,\frac{S_E^2}{4\pi^2}\:\exp\left(13.15 - \frac{15.8}{x^{2.9}}\right)
\times \exp\left(-S_E\right)
\end{equation}
with 
\begin{equation}
S_E\simeq\frac{f^4}{\Lambda^4}\sqrt{(1-x^2)\,(1-0.86 x^2)}\;\frac{4}{\pi} \left(\frac{12}{x}\right)^3\,,\qquad x = \frac{f\mu^3}{\Lambda^4}\,.
\end{equation}
The above expressions\footnote{The approximation~\eqref{eq:final_approximation} is valid as long as gravitational corrections to the tunneling rate are negligible (which we explicitly verified for the parameter combinations provided in this section).} allow us to directly determine parameter combinations $\mu$, $\Lambda$, $f$ which give rise to successful Chain EDE. As an example we choose
\begin{equation}\label{eq:parameters1}
\mu= 29.8\:\text{meV}\,,\quad
f= 13.9\:\text{meV}\,,\quad
\Lambda= 26.4\:\text{meV}\,,\quad
V_0 =(0.7\:\text{eV})^4\,,
\end{equation}
yielding an axion mass $m_\phi\sim\Lambda_0^2/f=50\:\text{meV}$. The initial field value is set to $\phi_0=0$ as mentioned previously. For the parameter choice above, the tunneling rate takes the value $\Gamma^{1/4} = 4\pc^{-1}$ corresponding to a lifetime $\tau_i=0.6\:\text{yr}$ per vacuum which remains constant along the chain.\footnote{We assume that there is no backreaction of the dark sector energy density $\rho_{\text{DS}}$ on the tunneling rate.} The energy density $\rho_{\text{EDE}}$ follows precisely the evolution depicted in Fig.~\ref{fig:edescenario} until the entire vacuum energy $\rho_\phi$ has been dissipated at $z\simeq 3300$. The number of phase transitions is $N=10^5$ such that CMB and LSS constraints (see Sec.~\ref{sec:cmbchain}) are easily satisfied.

The only problem of the tilted cosine model~\eqref{eq:potential} is that it lacks a mechanism to stop the axion once $\rho_\phi=0$, i.e.\ which prevents $\phi$ from further tunneling down the chain into the regime of negative vacuum energy. However, we remind the reader that the tunneling rate between two adjacent minima is exponentially sensitive to the parameters in the potential. Hence, small changes in the energy difference or barrier height between minima can quickly change the tunneling rate from fast to slow, i.e. prevent $\phi$ from further tunneling. 

We, therefore, now extend the tilted cosine model by a stopping mechanism for the axion. For this purpose we consider the potential
\begin{equation}\label{eq:relax}
 V = (M^2 - g_1 M \phi)\chi^2 - g_2 M^3 \phi + (\Lambda_0^4 +\Lambda_1^2 \chi^2)\cos\frac{\phi}{f} + \lambda\chi^4 + V_0\,,
\end{equation}
which has originally been motivated in the context of the relaxion mechanism~\cite{Graham:2015cka}.\footnote{We consider the non-QCD version of the relaxion mechanism, see Sec.\ III in~\cite{Graham:2015cka} and~\cite{Flacke:2016szy}.}

However, in contrast to the relaxion mechanism, we identify $\chi$ with a scalar field in the dark sector (rather than with the Higgs boson). The above potential has been argued to be radiatively stable since the breaking of the axionic shift symmetry is controlled by the (small) couplings $g_1\sim g_2$. We will, furthermore, assume that $M$ is much larger than the axion mass $m_\phi\sim\Lambda_0^2/f$.

\begin{figure}[htp]
\begin{center}
\includegraphics[width=0.5\textwidth]{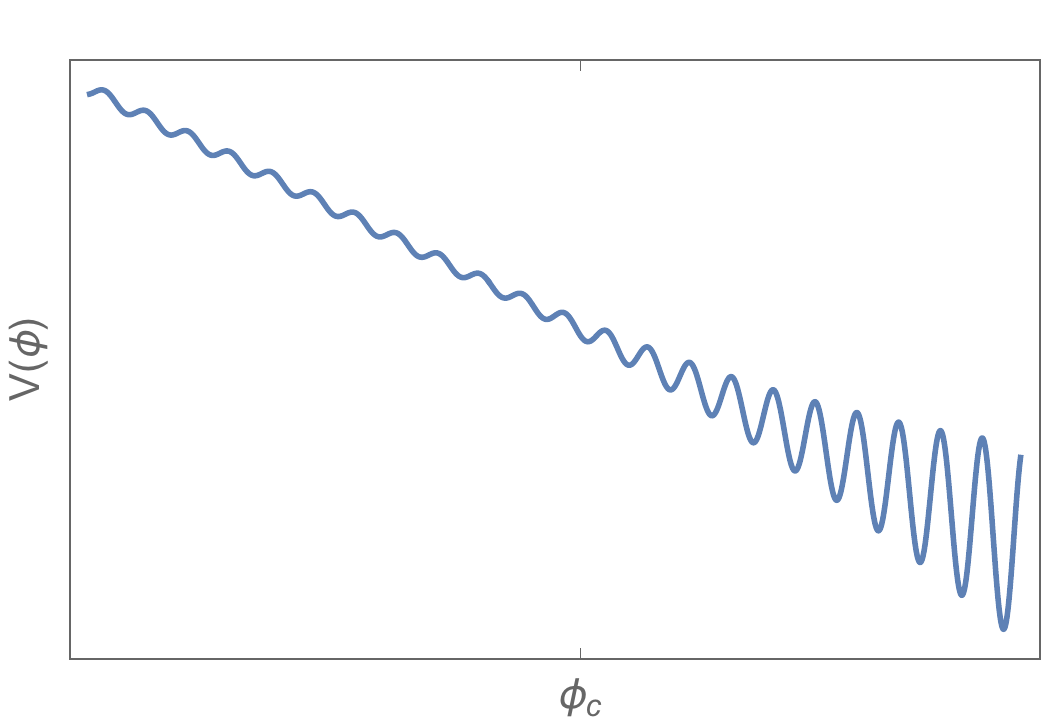}
\end{center}
\caption{Illustration of the potential~\eqref{eq:relax} in the axion direction with $\chi$ set to its respective minimum. The barrier height between minima remains constant as long as $\phi<\phi_c$, but increases quickly once the axion passes the critical value $\phi_c$. The axion tunneling rate between minima almost immediately switches from fast to slow at $\phi_c$.}
\label{fig:axionbarrier}
\end{figure}

Let us now look at the evolution of the two field-system starting from $\phi=0$. The field $\chi$ is initially stabilized at $\chi=0$ by the large mass term $M$ and can be integrated out. We thus obtain
\begin{equation}\label{eq:simplerelax}
 V = - g_2 M^3 \phi + \Lambda_0^4 \cos\frac{\phi}{f} + V_0\,,
\end{equation}
in the axion direction which agrees with~\eqref{eq:potential} if we identify $\mu\equiv g_2^{1/3} M$ and $\Lambda\equiv \Lambda_0$. The axion tunnels down the potential with the time spent in each vacuum remaining constant. However, once it reaches a field value $\phi_c\simeq M/g_1$, the squared mass of $\chi$ turns negative and $\chi$ gets displaced from the origin.\footnote{Notice that $\phi_c$ is only approximately given by $M/g_1$. This is because the term $\Lambda_1^2\chi^2 \cos\phi/f$ yields an additional subdominant mass term for $\chi$ which slightly shifts the transition.}  The quartic term stabilizes $\chi$ at a finite field value. As soon as $\chi\neq 0$, the term $\Lambda_1^2\chi^2 \cos\phi/f$ increases the barriers in the axion potential. Therefore, the tunneling time between vacua increases rapidly and becomes larger than the age of the universe shortly after the axion has passed $\phi_c$. In Fig.~\ref{fig:axionbarrier} we (schematically) depict the potential in the axion direction with $\chi$ set to its $\phi$-dependent minimum. 

In order to realize a successful EDE scenario, we consider the following parameter example,
\begin{align}\label{eq:parameters2}
M&=1.6\:\text{eV}\,,\quad
\Lambda_0= 26.4\:\text{meV}\,,\quad
\Lambda_1= 42.8\:\text{meV}\,,\quad
f= 13.9\:\text{meV}\,,\quad
C=(0.69\:\text{eV})^4\,,  \nonumber\\
\lambda&=0.002\,,\quad
g_1=1.8\times 10^{-4}\,,\quad
g_2=0.66\times 10^{-5}\,.
\end{align}
Starting from $\phi_0=0$ the axion undergoes $\sim 10^5$ tunnelings until it reaches the critical field value $\phi_c$. In the field range $\phi=[\phi_0,\,\phi_c]$, the axion follows exactly the dynamics of the tilted cosine model in Eq.~\eqref{eq:potential} (the parameters~\eqref{eq:parameters2} were chosen to reproduce~\eqref{eq:parameters1}). The lifetime of each vacuum remains constant at $\tau_i=0.6\:\text{yr}$. But once the axion passes $\phi_c$, the tunneling rate between vacua starts decreasing dramatically. For the specific example~\eqref{eq:parameters2}, only three more tunneling events occur after passing $\phi_c$ with corresponding lifetimes $\tau_i\simeq  1\:\text{yr}$, $60\:\text{yr}$ and $0.4\:\text{Myr}$. The next transition in the chain would already take $\sim 20\:\text{Gyr}$, i.e.\ longer than the age of the universe. One might worry that the handful of late transitions (at $\phi>\phi_c$) could spoil the cosmological evolution. However, this is not the case as $\rho_{\text{EDE}}$ remains strongly subdominant in the late universe which we explicitly verified for the example above. We can, hence, conclude that the relaxion mechanism~\eqref{eq:relax} provides a successful exit from the EDE epoch.

We want to emphasize, however, that Chain EDE does not necessarily require two scalar fields. Another model with only one field is suggested in the remainder of this paragraph.
A model building challenge for single-field realizations of Chain EDE consists in the prompt transition from rapid tunneling to a (meta)stable ground state. Most of the EDE must be dissipated around matter-radiation equality in order not to affect the late-time evolution of the universe. This could happen via a trigger mechanism like in Eq.\eqref{eq:relax}. However, another simple possibility is to consider a potential in which the barrier height between minima continuously decreases along the chain\footnote{Potentials of this type have e.g.\ been considered in the context of modulated natural inflation~\cite{Kappl:2015esy,Winkler:2019hkh}.}. The EDE field would tunnel quicker and quicker between minima until most of the EDE has decayed away. In the last stage the barriers in the potential become so shallow (or disappear entirely) that $\phi$ starts rolling. If the potential features a stable minimum at $V=0$, the EDE field would perform coherent oscillations around the minimum. Different from the oscillating scalar field EDE models discussed in Sec.~\ref{sec:oscillating} most of the EDE would, however, already be dissipated in the previous tunneling stage. The remaining subdominant EDE fraction would typically redshift at least as fast as matter during the oscillation stage. Given this fraction is sufficiently small compared to the initial EDE density, it should not significantly affect the further evolution of the universe (it would essentially manifest as a tiny fraction of the observed dark matter density). We leave a more dedicated analysis of this scenario and further Chain EDE realizations for future work.

\section{Chain Dark Energy}\label{sec:chainde}

In this section we suggest a new model for the Dark Energy (DE) that currently dominates the energy density of the Universe.
Again we imagine a chain of tunneling events. A scalar field starts somewhere up in the potential. This time, after the field successfully tunnels through a series of higher energy minima, it gets stuck in a low-energy false vacuum $\rho_{\text{DE}} \simeq (2\:\text{meV})^4$ with a lifetime longer than the current age of the Universe. The energy of this false vacuum could be responsible for the Dark Energy today. 

An enthralling possibility is that the same field is responsible for the EDE and the DE simultaneously. Comparing the EDE and DE energy densities, we have 
\begin{equation}
\frac{\rho_{\text{EDE}}}{\rho_{\text{DE}}}\simeq \left(\frac{0.7\:\text{eV}}{2\:\text{meV}}\right)^4 \simeq 10^{10}\,.
\end{equation}
This ratio can find a striking explanation within the EDE scenario and simply correspond to the number of phase transitions required to dissipate the EDE.

\begin{figure}[htp]
\begin{center}
\includegraphics[width=0.55\textwidth]{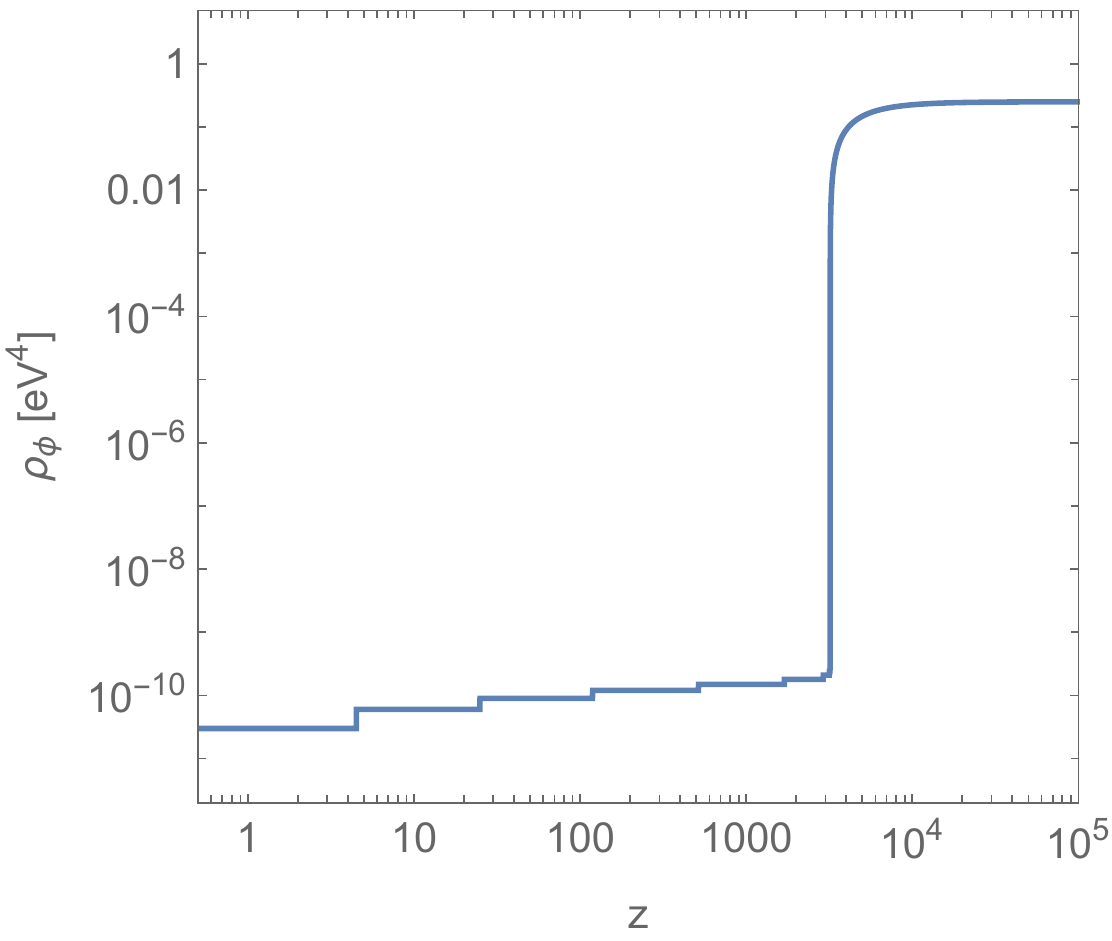}
\end{center}
\caption{Illustration of the scenario, where the same scalar field $\phi$ accounts for the EDE and the DE. Depicted is the vacuum energy stored in $\phi$ as a function of redshift. At large redshift $\rho_\phi$ remains approximately constant. But around matter-radiation equality $\rho_\phi$ decreases quickly by fast tunneling along the chain of vacua. The energy is dumped into dark radiation or anisotropic stress (not shown in the picture). Once most of $\rho_\phi$ has been dissipated, the tunneling rate becomes small and only a few more vacuum transitions occur within the lifetime of the universe (shown as the steps in the above figure). The energy density in the final vacuum in which $\phi$ settles until today corresponds to the DE of our present universe.}
\label{fig:DEEDE}
\end{figure}

For illustration let us consider the Chain EDE model described in the previous section (see Eq.~\eqref{eq:relax}): the EDE field tunnels quickly through a large number of vacua until matter-radiation equality. But once most of the EDE has been dissipated, the lifetime of individual vacua blows up and only a few more tunneling events occur. If we set the initial EDE density to $(0.7\:\text{eV})^4$ and require $N\simeq 10^{10}$ phase transitions, the energy difference between individual vacua comes out as
\begin{equation}
\Delta V = (2\:\text{meV})^4\,.
\end{equation}
If the EDE field settles in the lowest de Sitter minimum, the corresponding energy density is of $\mathcal{O}(\Delta V)$. Hence, it could naturally account for the DE which dominates our universe today. Only after a time longer than the age of the universe, the EDE field would ultimately tunnel into the next minimum along the chain with negative energy. Far in the future our observable universe would then end in a big crunch. For illustration we depict the evolution of the energy density stored in the EDE field as a function of redshift in Fig.~\ref{fig:DEEDE}.

Our chain model relates the cosmological constant to the parameters in the axion potential. Furthermore, it successfully establishes a connection between the EDE and DE energy densities in terms of the number of phase transitions. However, as yet our model does not provide a solution of the cosmological constant problem. Within the mechanism~\eqref{eq:relax} the fact that the axion stops tunneling at the right moment in time (i.e.\ when the vacuum energy is small) is a coincidence and relies on a parameter choice. A full solution to the cosmological constant problem without fine-tuning would require a dynamical reason for the axion to stop in the desired minimum.

An intriguing idea in this direction has been formulated by Abbott in 1984~\cite{Abbott:1984qf} who suggested a solution to the cosmological constant problem in terms of a tunneling field -- similar to our Chain Dark Energy proposal. However, he relies on gravitational corrections to the tunneling in order to dynamically stop the tunneling field in a vacuum with small energy density. Abbott's proposal fails since it requires an extremely flat potential of the tunneling field (in order to make gravitational corrections important) which renders the phase transitions far too slow to relaxate the vacuum energy within the age of the universe. Nevertheless, it would be very interesting to explore, whether we can employ a dynamical stopping mechanism for the axion in our chain (E)DE scenario which relies on gravity. For example one could try to extend the Abbott mechanism by a non-minimal coupling of the the tunneling field to gravity in order to increase gravitational corrections. We leave further investigation of the cosmological constant problem within our chain models for future work.

\section{Conclusion and Discussion}

We have suggested Chain Early Dark Energy as a solution to two problems in cosmology,
the Hubble tension and today's small value of the Dark Energy.  
The original idea of EDE was proposed~\cite{Karwal:2016vyq,Poulin:2018cxd} to resolve
the apparent discrepancy between local measurements of the Hubble constant $H_0\simeq 74 \km\,\s^{-1}\Mpc^{-1}$~\cite{Riess:2018byc,Riess:2019cxk} 
and $H_0\simeq 67 \km\,\s^{-1}\Mpc^{-1}$ inferred from the CMB~\cite{Aghanim:2018eyx} by altering the expansion history of the Universe right around the epoch of matter-radiation equality.
The original EDE model employed a scalar field oscillating in a potential.

In Chain EDE, the Universe instead undergoes a series of first order phase transitions, starting at a vacuum with energy density $\rho_{\text{EDE}}=\mathcal{O}( \text{eV}^4)$, and tunneling down through a chain of metastable minima with decreasing energy. For sufficiently many transitions, the discreteness of individual vacua can be neglected and $\rho_{\text{EDE}}$ decreases quasi-linearly with time. As can be seen in Fig.~\ref{fig:linear}, the linear decrease looks almost like a step function in a log-log-plot. Such a rapid drop of vacuum energy produces a narrow peak in the energy fraction $\rho_{\text{EDE}}/ \rho_{\text{tot}}$ as a function of $\log t$ (or equivalently as a function of $\log z$, see Fig.~\ref{fig:edescenario}). With the appropriate choice of two parameters -- the initial EDE density $V_0$ and the summed lifetime of vacua along the chain $N/\Gamma^{1/4}$ -- the EDE fraction of the Universe's energy density is negligible at the beginning, raises to $\sim 10\%$ around matter-radiation equality and then decays away quickly, as shown in Figure~\ref{fig:edescenario}. This is exactly the required behavior to resolve the Hubble tension: the additional contribution to the energy density of the Universe prior to recombination reduces the sound horizon at matter-radiation decoupling and leads to a larger value of $H_0$ derived from the CMB.

In principle the bubbles formed from first order phase transitions in the early Universe  
could leave dangerous imprints in the CMB or affect LSS. However, we have shown that if the lifetime of individual vacua does not exceed a few years, the scale of the anisotropies is below the resolution of current experiments, i.e.\ all observational constraints are satisfied. Since the solution to the Hubble tension requires the EDE field to stay around for about $50000$ years (until matter-radiation equality) the anisotropy constraints impose $N\gtrsim 10^4$ phase transitions.
Further, since Chain EDE has essentially the same background evolution as previously studied EDE models which were shown to resolve the Hubble tension while satisfying all cosmological constraints~\cite{Smith:2019ihp}, we expect the same to be true for Chain EDE.

We have provided a concrete model of chain EDE  which easily achieves the required number of transitions and which carries strong motivation from axion physics. The model employs a scalar field in a quasi-periodic potential (a tilted cosine). The scalar field tunnels from vacuum to vacuum at (nearly) constant rate until it reaches a critical field value, where it is stopped almost instantly, e.g. due to the backreaction from a second scalar field (employing the relaxion mechanism). Further pathways to successful Chain EDE models with only a single scalar field are also discussed.

Interestingly, the energy difference between vacua can be of the same size as the Dark Energy density of the present universe (meV scale). This offers the exciting prospect to explain EDE and DE by the same scalar field. The tunneling field would be subdominant to ordinary matter and radiation throughout its evolution, until it gets trapped in the last minimum of the chain before reaching zero/negative energy. If this minimum has a lifetime longer than the age of the universe, the remaining vacuum energy of the scalar field would produce today's Dark Energy.

We end with speculation about Recurrent Chain Dark Energy.
Most exciting of all would be a Chain vacuum energy model that could explain all epochs in the history of the Universe, where the vacuum energy dominates or becomes significant: inflation, EDE, today's Dark Energy (and perhaps others we do not yet even know about).  We imagine a Chain potential in which a field tunnels through a series of ever lower minima.  In this Recurrent Model, the vacuum energy is initially the dominant energy density in a Chain Inflationary epoch; after that the vacuum would be mostly subdominant to radiation and matter, but occasionally raises its head to a large enough value to affect
the Universe evolution.  It becomes important at $z\sim 3000$ at the level of  10\% of the total energy density to provide the EDE that can resolve the Hubble tension, and it is dominant again today as the origin of the Dark Energy.   
Recurrent Chain Dark Energy would most likely require multiple scales in the potential (rather than e.g.\ a single tilted cosine).  The difficulty of this idea is that
 inflation must reheat to the Standard Model (SM), but EDE must reheat to a dark sector (suggesting no direct coupling to visible matter) in order to avoid unacceptable modifications to the CMB. It would be interesting to look for a successful model of Recurrent Dark Energy that avoids this problem, e.g. inflation producing
very massive DS particles that later decay to the SM particles, but which are too massive to be created in the EDE epoch.

\section*{Acknowledgments}
We thank Martina Gerbino, Jon Gudmundsson,
Dragan Huterer, Massimiliano Lattanzi, and Sunny Vagnozzi for helping us understand bounds on anisotropies.
KF thanks Matt Johnson and Jim Liu for extremely helpful discussions about the cosmological constant problem.
We also thank Jason Pollack and Mike Boylan-Kolchinski for our conversations about the paper.
K.F.\ is Jeff \& Gail Kodosky Endowed Chair in Physics at the University of Texas at Austin, and K.F.\ and M.W.\ are grateful for support via this Chair.  K.F.\ and M.W.\ acknowledge support by the Swedish Research Council (Contract No. 638-2013-8993).

\pagebreak
\bibliography{ede}
\bibliographystyle{h-physrev.bst}

\end{document}